# MINIATURIZED FLUORESCENCE EXCITATION PLATFORM WITH OPTICAL FIBER FOR BIO-DETECTION CHIPS

*Hsiharng Yang [1, 2] and Chung-Tze Lee [1]*

[1]Institute of Precision Engineering, National Chung Hsing University, Taichung, Taiwan 402
[2]Center of Nanoscience and Nanotechnology, National Chung Hsing University, Taichung, Taiwan 402

**ABSTRACT**

This paper presents a new research study on the platform fabrication of fluorescence bio-detection chip with an optical fiber transmission. Anisotropic wet etching on (100) silicon wafers to fabrication V-groove for optical fiber alignment and micro-mirror were included. Combing with anodic bonding technique to adhere glass, silicon structure and optical fiber for a fluorescence excitation platform was completed. In this study, the etching solution 40% KOH was used to study the parameters effect. The results show that working temperature is the main parameter to significantly effect the etch rate. The anisotropic etching resulted 54.7° reflective mirrors and its reflectivity for optical beam were also examined. The surface roughness of the micro-mirror is Ra 4.1 nm measured using AFM, it provides excellent optical reflection. The incident light and beam profiles were also examined for further study. This study can show this micro-platform adaptable for fluorescence bio-detection.

## 1. INTRODUCTION

The progressive development of biotechnology and micro-optical system technology trends to a mature stage for practical applications. By adding integration packaging methods, many bio-detection systems are able to realize "lab on a chip". Lab on a chip provides some advantages on bio-experiments such as simple, fast multi-processing and high accuracy [1]. There are many bio-detection systems are widely studied such as electrochemical method, refractive metrology, fluorescent detection, mass spectrometry. The fluorescent detection is the most common technique for bio-detection. It uses a certain wavelength and high energy beam such as a laser to excite unknown material characters with fluorescent particles. By reading the excited fluorescent intensity, the unknown concentration of fluorescent molecules per unit volume can be obtained. In practical manipulation, it is a high sensitive and selective method. The conventional bio-fluorescent detection uses a heavy and expensive instrumentation. They are high detection cost and not portable. Miniaturizing bio-detections are proposed by many researchers to realize the detection in microsystems.

Warren et al. proposed the integration of light source, micro-fluidic channels, filter, and detection on a chip [2]. It's a quite highly integrated fluorescent microsystem, but the chip is hard to clean or cannot repeat to use. Hubnera et al. proposed a transportable miniaturized fiber-pigtailed measurement system for fluorescence spectroscopy in microfluidic channels [3]. Excitation light was launched into the microchannel by a single-mode fiber guide on top of the microchannel. A fiber terminated waveguide was used to collect fluorescence light. Optical losses between waveguides and fibers may result in low power efficiency. Camou et al. tried to use thick film process and PDMS replication method to fix a fiber and microlens [4]. It made the fiber easy to be aligned and fixed excitation light as well as focused light to fluidic microchannel. The detector can be simple to place on top or bottom for detection. One mask pattern for fabricating PDMS mold is needed. It is simple fabrication process and low cost and becomes popular applications.

Roult et al. used double sides of microlens for coupling and focus light [5]. Micochannels were made using isotropic etching glass for fluid. Microlens array was fabricated using thermal reflow on the other glass. Two glasses were bonded as a chip. Excitation light was a laser inclined 45˚ through a microlens to excite fluorescent light. The other microlens was used to collect the fluorescent light and guide to the detector. Microlenses were widely applied in fluidic microchannel fluorescent detection [6, 7]. Using LEDs as the excitation light is also a new progress [8, 9]. LEDs were directly embedded onto the chip, they eliminated to use fibers. The integrated microfluidc chips directly combined microlens and fluidic microchannels. They are quite low cost compared to previous systems and disposable from other contaminations.

An excitation light platform for fluorescent light detection using disposable microfluidic chips is proposed. A laser excitation light guided by optical fiber is aligned to silicon V-groove and reflected by a micromirror in V-groove. The disposable microfluidic chip can be placed on top of the platform. The detector can read out the fluorescent light. It should provide a high detection light output in the platform. Thus, the miniaturized detection system is portable and suitable for disposable microfluidic chips.

## 2. EXPERIMENTAL METHODS
### 2.1 Detection method
The fluorescent detection method is a promised technique for miniaturized bio-detection systems. The conventional instrumentation relies on a fluorescent microscope to detection the lower limit concentration of bio-fluid. It's quite practical, but high instrumentation cost and low mobility are their disadvantages. The miniaturized system can overcome those setbacks and provides high accuracy and mobile handling. In this miniaturized bio-detection platform, the excitation light is guided by an optical fiber through a reflective mirror and transmitted to a microchannel with fluorescent labels. By using this platform, a closely spaced microfluidic channel and its array can be scanned.

### 2.2 Design of V-groove etching
The silicon V-groove is used to auto-align an optical fiber. It is a sophisticated technique by using single crystalline of silicon anisotropic etching [10, 11]. There is a unique angle of 54.74° between (100) and (111) lattice planes of silicon crystalline. Anisotropic etching (100) silicon may result in a symmetric V-groove, it is a perfect alignment profile for a round article such as a bare optical fiber. It will need an opening channel with **w** in width and **d** in depth. The related geometry of V-groove and an optical fiber is sketched in Figure 2. Where **r** is the radius of a bare optical fiber, the actual dimension is 62.5 μm. θ is a fixed angle 54.74° and **x** is a variable number dependent on the depth **d**. The mathematical relationship between each parameter is described below. Equation (1) shows the opening width **w** related to the etching depth **d** and angle θ. The required etching depth **d** is also related to the radius of the optical fiber **r** as shown in Equation (2). After input **r** = 62.5 μm, the minimum etching depth **d** is equal to 172 μm and **w** is equal to 250 μm. An optical fiber can be inserted into the silicon V-groove platform.

$$w = 2d \cot \theta \qquad (1)$$

$$\cos\theta = \frac{r}{r+x} \;\Rightarrow\; x = r(1-\cos\theta)\sec\theta \qquad (2)$$

$$d = 2r + x \;\Rightarrow\; d = 2r + r(1-\cos\theta)\sec\theta$$

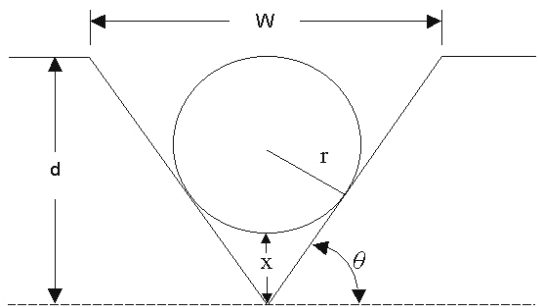

Figure 2. Schematic of V-groove geometry related to an optical fiber.

### 2.3 Processing of V-groove platform
Once the related dimensions of V-groove are determined, a photomask with the designed pattern is fabricated. Figure 3 illustrates every step to fabricate the V-groove platform. A 1.0 μm thick $SiO_2$ film (oxidation) was grown using wet oxidization at 1100 °C on a one-sided polished <100> wafer shown in step 1. The bulk etching area was patterned on the silicon wafer as shown in step 2 to 3. Wet oxidization is regarded as the etching mask. The oxidation removal was done using BOE (buffered oxide etchant) in step 4. Step 5 shows the resist removal using acetone. The V-groove etching was done using aqueous KOH in step 6. The whole wafer was immersed into 40 wt% aqueous KOH solution at 70 °C with magnetic stirring. Various experimental parameters including temperature affect etching rate and depth control were studied. Then the dioxide etch stop layer is then etched with BOE solution in step 7. Aluminum coating was done by using thermal evaporation in step 8. Figure 4 is a SEM micrograph of the finished V-groove in silicon. It is an over etching result, the channel width is larger than the desired size. However, the optical fiber is still can be aligned and inserted as shown in Figure 5.

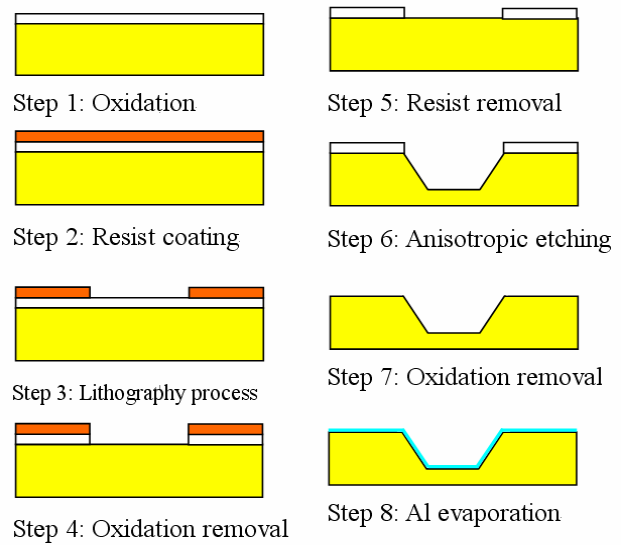

Figure 3. Fabrication process of V-groove etching for the detection platform.

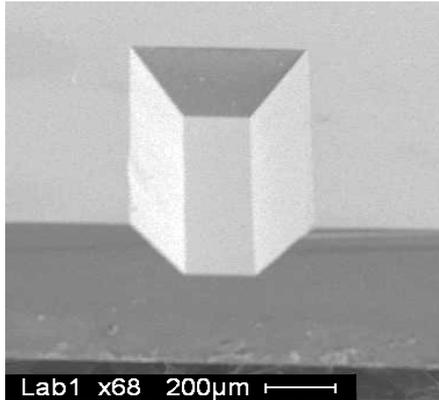

Figure 4. SEM micrograph of V-groove in silicon.

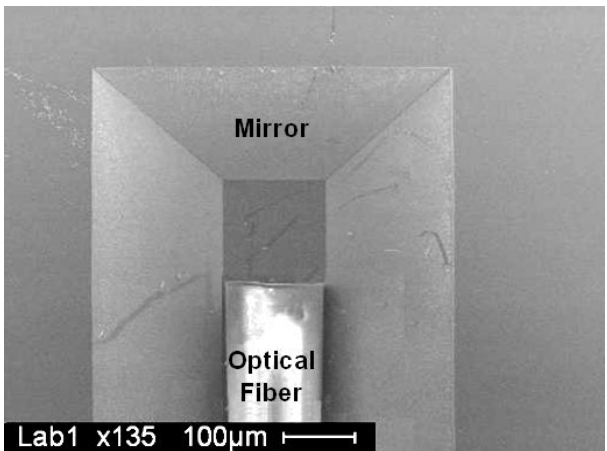

Figure 5. SEM micrograph of optical fiber inserted in silicon V-groove.

### 2.3 System assembly

Silicon V-groove was used as a fiber holding platform. A Pyrex #7740 corning glass with thickness 500 $\mu$m was to cover the V-groove platform with the fiber. Anodic bonding process was used to bonding the silicon chip and glass. Before the bonding, the optical fiber was inserted into V-groove and adhered by UV curable glue. The anodic bonding equipment included high voltage DC power supply, hot plate, and counterpoise to increase weight for bonding tightness. The positive electrode was connected to a graphite sheet to provide a uniform electrical field. The anodic bonding steps were listed below. DI water was firstly used to clean the chip surface and then spray using nitrogen flow. Followed to dip in a clean solution ($H_2SO_4$ : 300 ml、$H_2O_2$ : 100 ml) at temperature 80℃ for 20 minutes. Rinsed by DI water for 1 minute and sprayed by nitrogen flow. The graphite as an electrode in the anodic machine was heated to 500℃. The silicon chip and Pyrex glass were placed onto the graphite as anodic. The counterpoise was connected to cathode. The input voltage was 700 V. Two weights of counterpoises were used for bonding process. The bonding time was 40 minutes for counterpoise 48 g and 20 minutes for 80 g. Both bonding conditions were successful to joint the silicon chip and glass. The assembly result is shown in Figure 6. However, the counterpoise weight related to bonding time was not investigated. It could be studied in the future.

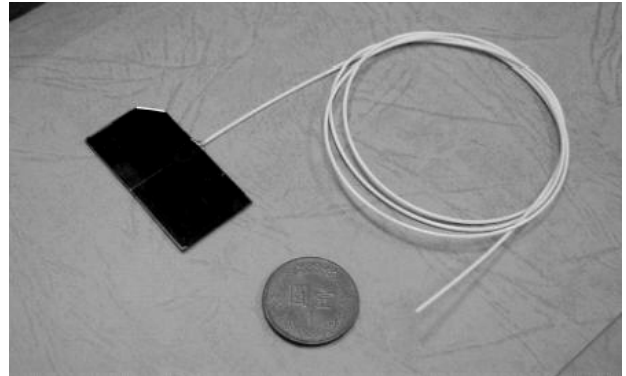

Figure 6. System integration of the bio-detection platform.

### 3. RESULTS AND DISCUSSION

The present result shows some basic microfabrication techniques for the bio-detection platform manufacturing. Silicon anisotropic etching was used to make V-groove channels and reflective mirrors. Surface quality of the reflective mirror and optical performance of the platform are discussed here.

### 3.1 Anisotropic etching for V-groove fabrication

Silicon anisotropic etching (100) wafer was operated using 40 wt% KOH solutions. Precise control etching rate helps to determine the etching time. Different etching temperatures including 40, 60, 70, 80, and 90℃ were studied for their related etching rate as shown in Figure 7. The etching rates are ranged from 0.25 to 2.0 μm per minute dependent on the etching temperature. Each etching depth was measured by a 3D surface profilometer, then divided by the etching time for its etching rate.

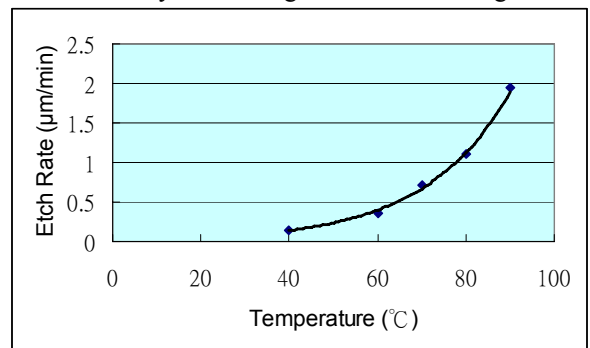

Figure 7. Silicon (100) etching rate related to temperature.

### 3.2 Surface roughness of V-groove

Optical reflective mirror is always required a smooth surface for high reflectivity. Figure 8(a) shows the surface roughness measurement result of the (111) silicon lattice plane by using AFM. The average surface roughness Ra 4.2 nm was measured on an area of 2 by 2 μm$^2$. On the other hand, average surface roughness Ra 20.9 nm was measured on (100) lattice plane surface as shown in Figure 8(b). However, the reflective mirror being used is (111) lattice plane. Basic physical property on surface roughness is achieved, but it has to be examined by optical beam measurement for its reflectivity.

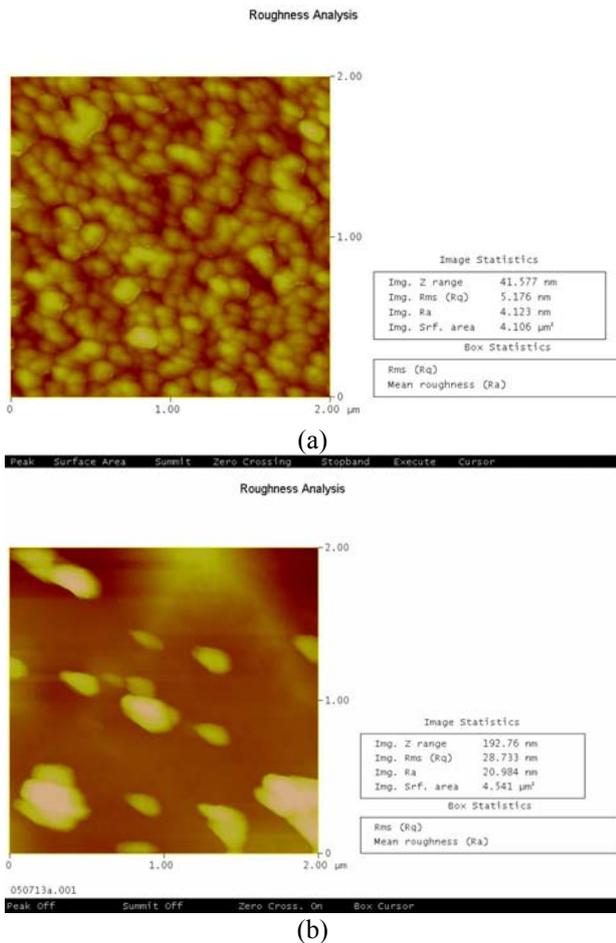

(a)

(b)

Figure 8. Surface roughness of V-groove surface using AFM measurement; (a) (111) lattice plane and (b) (100) lattice plane.

### 3.3 Optical measurement of the reflective micro-mirror

The optical measurement setup for the miniaturized bio-detection platform is illustrated in Figure 9. A laser source with wavelength 632 nm was used as an input for the optical fiber. The fiber was aligned in V-groove and guided the laser beam. The (111) lattice plane is a reflective mirror to bend the laser beam. A photo detector was placed on the top to detect the output and reading from the power meter. Two kinds of samples were used to test the optical mirror reflectivity; the mirror surface without Al coating and with Al coating as shown in Table 1. Samples 1 to 3 without Al coating were measured to average 31% reflectivity. Samples 4 to 6 with Al coating (100 nm thick) showed average 70% reflectivity. The visual optical shapes of the excitation light captured by the detector are shown in Figure 10. They are a comparison between output from the fiber (Fig. 10(a)) and output from the reflective mirror. The excitation light is scattered after the reflection. It indicated that a microlens to refocus the excitation light after the reflection may need to be developed in the future.

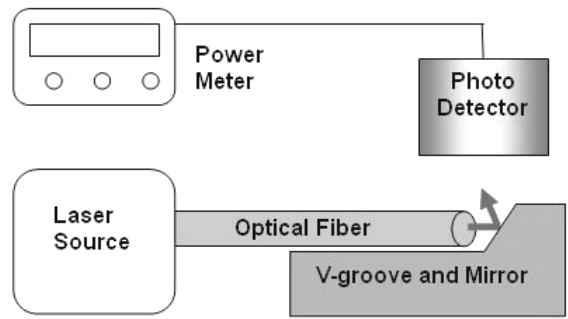

Figure 9. Optical measurement setup for the detection platform.

Table 1. Optical reflectivity measurement.

| Power<br>Surface | | Input<br>P1 (μW) | Output<br>P2 (μW) | Reflectivity<br>(P2/P1) |
|---|---|---|---|---|
| Without Al coating | Sample 1 | 0.9735 | 0.2864 | 29% |
| | Sample 2 | 0.9132 | 0.3054 | 33% |
| | Sample 3 | 0.9346 | 0.3011 | 32% |
| With Al coating (100 nm thick) | Sample 4 | 0.9234 | 0.6602 | 71% |
| | Sample 5 | 1.0191 | 0.7094 | 69% |
| | Sample 6 | 0.9531 | 0.6871 | 72% |

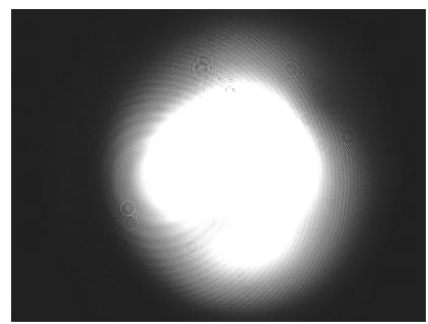

(a)

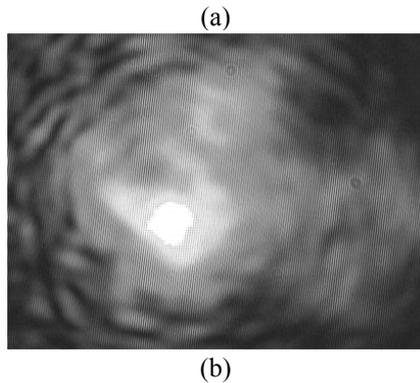

(b)

Figure 10. Optical shape measurement result of the excitation light; (a) output from optical fiber and (b) output from the reflective micro-mirror.

## 4. CONCLUSION

The experiments were started at silicon etching process to fabricate a V-groove for holding an optical fiber. Various working temperatures were studied from the range of 40 to 90℃ by using the etching solution 40% KOH. The highest etching rate was 2 μm per minute at temperature 90℃. A V-groove with width 250 μm and depth 172 μm in (100) silicon substrate was fabricated. One end in the groove with 54.7° was also made to be a reflective mirror. The primitive reflective mirror was tested by a fiber laser with wavelength 632 nm. Only average 31% reflectivity was obtained from three samples. To improve the mirror reflectivity, thermal evaporation aluminum (Al) 100 nm film on the mirror can reach 70% reflectivity. They were examined by using an optical image system. The Al coating reduced light scattering from the mirror surface. AFM surface roughness examination showed the micro-mirror only Ra 4.1 nm after Al coating. With a high reflective micro-mirror on the end of the groove, the optical fiber was placed in silicon substrate. The fluorescent detection platform was covered with a Pyrex #7740 glass by anodic bonding. In connect with the fiber laser and capillary electrophoresis microchannels, the platform can be used to detect labeled fluorescent particles in the future.

## 6. ACKNOWLEDGEMENT

This work was supported by the National Science Council (series no. NSC94-2212-E-005-016) of Taiwan, R.O.C.

## 7. References

[1] Chabinyc, M. L., Chiu, D. T., Mcdinald, J. C., Stroock, A. D., Christian, J. F., "An integrated fluorescence detection system in poly(dimethylsiloxane) for microfluidic applications," Anal. Chem, Vol.73, pp. 4491-4498, 2001.

[2] Warren, M. E., Wendt, J. R., Sweatt, W. C., Bailey, C. G., Matzke, C. M., Asbill, D. W. and Samora, S., "VCSEL-based micro-optical system," IEEE, 2, pp. 407-408, 1998.

[3] Hubnera, J., Mogensen, B., Jorgensen, M., Friis, P. Telleman, P. and Kutter, J. P., "Integrated optical measurement system for fluorescence spectroscopy in microfluidic channels," Review of Scientific Instruments, 72(1), pp. 229-233, 2001.

[4] Camou, S., Gouy, J. P., Fujita, H. and Fujii, T., "Integrated 2-D optical lenses designed in PDMS layer to improve fluorescence spectroscopy using optical fibers," IEEE, Vol.1, pp. 187 – 191, 2002.

[5] Roulet, J. C., Volkel, R., Herzig, H. P., Verpoorte, E., Rooij, N. F. and Dandliker, R., "Microlens systems for fluorescence detection in chemical microsystems," Opt. Eng., 40(5), pp. 814–821. 2001.

[6] Adams, M. L., Enzelberger, M., Quake, S. and Scherer, A., "Microfluidic integration on detector arrays for absorption and fluorescence micro-spectrometer," Sensor and actuators A, 104, pp. 25-31, 2003.

[7] Jeong, K. H. and Lee, L. P., "A new method of increasing numerical aperture of microlens for biophotonic MEMS," 2nd Annual International IEEE-EMBS Special Topic Conference, pp. 380-383, 2002.

[8] Chediak, J. A., Luoa, Z., Seo, J., Cheungc, N., Lee L. P. and Sandse, T. D., " Heterogeneous integration of CdS filters with GaN LEDs for fluorescence detection microsystems," Sensors and Actuators A, 111(1), pp. 1-7, 2004.

[9] Seo, J. and Lee, L. P.,"Disposable integrated microfluidics with self-aligned planar microlenses," Sensors and Actuators B, 99, pp. 615–622, 2004.

[10] Kovacs, T. A., Maluf, N. I. and Petersen, K. E. "Bulk micromachining of silicon," Proceedings of The IEEE, 86(8), pp. 1536-1551, 1998.

[11] Marxer, C., Thio, C., Gretillat, M., Rooij, N. F., Battig, R., Anthamatten, O., Valk, B. and Vogel, P., "Vertical mirrors fabricated by deep reactive ion etching for fiber-optic switching applications," Journal of Microelectromechanical Systems, 6(3), pp. 277-258, 1997.